\documentclass{amsart}
\usepackage{graphicx,amssymb}
\def\dt{\partial_t}

\def\d{\mathrm{d}}

\def\dbold{\mbox{\boldmath${\partial}$}}

\newcommand{\Real}{\mathrm{Re}}
\newcommand{\eps}{\varepsilon}

\newcommand{\mbold}[1]{\mbox{\boldmath${#1}$}}
\def\beq{\begin{eqnarray}}
\def\eeq{\end{eqnarray}}
\begin{document}

\title{Inhomogeneous perturbations of plane-wave spacetimes}
\author{Sigbj{\o}rn Hervik \& Alan Coley}%
\address{Department of Mathematics \& Statistics, Dalhousie University,
Halifax, Nova Scotia,
Canada B3H 3J5}%
\email{herviks@mathstat.dal.ca, aac@mathstat.dal.ca}%

\subjclass{}%
\keywords{}%

\date{\today}%
\begin{abstract}
Recently it was shown that the exact cosmological solutions known as the vacuum plane-wave solutions are late-time attractors for an open set of the spatially homogeneous Bianchi universes containing a non-inflationary $\gamma$-law perfect fluid. In this paper we study inhomogeneous perturbations of these plane-wave spacetimes. By using expansion-normalised scale-invariant variables we show that these solutions are unstable to generic inhomogeneous perturbations. The crucial observation for establishing this result is a divergence of the expansion-normalised frame variables which ultimately leads to unstable modes.  
\end{abstract}
\maketitle
\section{Introduction}
Since the advent of general relativity researchers have sought physical relevant solutions to the
Einstein field equations.  However, due to the complicated nature of
the field equations symmetries are 
usually imposed in order to make the field equations more tractable; for example, the
spatially homogeneous Bianchi models in $(3+1)$-dimensional cosmology are of particular interest
\cite{bianchi,EM,WE,DS2}.  One of the most fascinating and interesting solutions to these field
equations are the so-called \emph{vacuum plane-waves}.  These solutions have some extraordinary
physical and mathematical properties which have drawn the attention of a broad spectrum of
researchers.  For the Bianchi models these plane-waves play a particular role; they are late-time
attractors for non-inflationary non-compact perfect fluid cosmologies of Bianchi class
B.

The spatially homogeneous (SH) Bianchi class B models contain  types IV, V, VI$_h$ and VII$_h$.
Bianchi models are referred to as orthogonal cosmologies when the
fluid flow is orthogonal to the group orbits. Otherwise they
represent `tilted' models. The types VI$_h$ and VII$_h$ are of
non-zero measure in the space of all spatially homogeneous
models and are thus what  might be called `general' type. The type  VII$_h$ models are of particular physical interest
because they contain open Friedmann-Robertson-Walker (FRW) models.
It is known that in general relativity all orthogonal class B perfect fluid models,
with equation of state parameter $\gamma \le
2$, expand indefinitely into the future and an open set of models
are asymptotic in the future
to a vacuum plane wave (self-similar) solution \cite{Siklos,hw}. 
In the stiff fluid case, apart from a set of measure zero,
the asymptotic form of the general type VII$_h$ stiff
perfect fluid solution corresponds to the vacuum type VII$_h$ plane wave solution first derived by Doroshkevich, Lukash
and Novikov~\cite{dln}. Orthogonal models are special; it is
therefore important  to consider the effect that tilt has on the
asymptotic behaviour of the class B models.

Until recently, only partial results regarding tilted Bianchi models were known \cite{BS,BHtilted}.
However, a detailed analysis of the tilted Bianchi models shows that many of the vacuum plane waves
are stable with regards to more general tilted matter \cite{CH,HHC}.  This analysis includes keeping
track of the asymptotic tilt velocity of the perfect fluid which, in
general, can have different kinds of
complicated asymptotic behaviour; e.g.,  attracting points, closed orbits and even attracting tori.
Notwithstanding this complex behaviour of the tilt velocity, the plane-wave geometry are
asymptotically stable for types IV, VII$_h$, and an open set of type VI$_{h}$ models.  Since the
types VII$_h$ and VI$_h$ are among the most general Bianchi models, we can draw a remarkable
conclusion:  \emph{In the class of all spatially non-compact Bianchi universes with a
non-inflationary $\gamma$-law perfect fluid (not necessary co-moving), there exists an open set of
universes asymptoting towards the vacuum plane-waves at late times}.

As stressed in \cite{WE,DS2}, an important mathematical
link exists between the various classes in the {\it state space}
hierarchy. The physical state of a cosmological model at an
instant of time is represented by a point in the state space,
which is finite dimensional for SH models and infinite dimensional
otherwise. This structure opens the possibility that the evolution
of a model in one class may be approximated, over some time
interval, by a model in a more special class. Thus it is plausible
that understanding the dynamics at one level of complexity,  such
as for example the spatially homogeneous Bianchi type VI$_h$  and VII$_h$
models, will shed light on the possible dynamical behaviour at a
higher level, such as special classes of inhomogeneous
cosmological models.

To investigate this further, in this paper we will consider
inhomogeneous perturbations of the Bianchi type VII$_h$ plane-wave
spacetimes. In particular, we will determine whether the plane-wave
solutions are future attractors for models of non-zero measure in
general relativity. However, our analysis indicates that this is not
so; the models are unstable with respect to inhomogeneous
perturbations. The critical unstable modes we find correspond to
inhomogeneous perturbations orthogonal to the propagation of the
gravitational wave. For these plane-wave spacetimes the direction
along the propagation of the wave expands linearly in cosmological
time, while the orthogonal directions expands more slower. The
instability of these models stems from this anisotropic expansion; the
slower expanding directions do not expand rapidly enough to prevent
the forming of inhomogeneities. 

In order to  understand this instability, it is illustrative to
compare the expansion with a flat FRW model. The flat FRW model
expands as $a(t)\propto t^p$, where $t$ is the cosmological
time. Inflation occurs for $p>1$, while $p<1$ leads to instability with regards to inhomogeneous perturbations  \cite{WE}. For the plane-wave spacetimes, one direction expands with $p=1$ (the threshold value), while the others expand with $p<1$. Hence, we expect instability in two of the directions (the third is a marginal case). 

\section{Inhomogeneous cosmologies} 
Recently, a formalism for studying the dynamics of inhomogeneous
cosmologies ($G_0$ models) by employing expansion-normalised scale-invariant variables was developed \cite{UEWE}. This formalism uses orthonormal frames to write the Einstein field equations as an autonomous system of evolution equations and constraints.

The orthonormal frame $\{ {\bf e}_0,{\bf e}_{\alpha}\}$ can be expressed in local coordinates as 
\beq
{\bf e}_0=\frac{1}{N}(\dt-N^i\partial_i), \quad {\bf e}_{\alpha}=e_{\alpha}^{~i}\partial_i. 
\eeq
We introduce the Hubble-normalised frame as follows: 
\beq
\dbold_0=\frac{1}{H}{\bf e}_0, \quad \dbold_{\alpha}=\frac{1}{H}{\bf e}_\alpha. 
\eeq
It is also necessary to introduce the deceleration parameter $q$ and the spatial Hubble gradient $r_{\alpha}$:
\beq 
q+1\equiv-\frac{1}{H}\dbold_0H, \quad r_{\alpha} \equiv-\frac{1}{H}\dbold_\alpha H.
\eeq
For the purposes of this paper, it is useful to introduce the separable volume gauge: 
\beq
N=H^{-1}, \quad N^i=0, \quad \dot{U}_{\alpha}=r_{\alpha}. \label{temporalgauge}
\eeq
This enables us to write
\beq
\dbold_0=\dt, \quad \dbold_{\alpha}=E_{\alpha}^{~i}\partial_i,
\eeq
where $E_{\alpha}^{~i}\equiv e_{\alpha}^{~i}/H$.

The Hubble-normalised state vector for $G_0$ cosmologies with a $\gamma$-law perfect fluid can then be given by 
\beq
{\bf X}=\left[E_{\alpha}^{~i},~r_{\alpha},~\Sigma_{\alpha\beta},~A^{\alpha},~N_{\alpha\beta},~\Omega,~v^{\alpha}\right]^T,
\eeq 
where  $\Sigma_{\alpha\beta}$ is the expansion-normalised shear, $A^{\alpha}$ and $N_{\alpha\beta}$ are the connection variables, and $\Omega$ and $v^{\alpha}$ are the energy density and the tilt of the matter, respectively. The evolution equations and the constraints can now be written down in terms of ${\bf X}$ (see \cite{UEWE}). The equations can be written in the form
\beq
\dt {\bf X}&=& {\bf F}({\bf X},\partial_i{\bf X},\partial_i\partial_j{\bf X}),\label{eq:evol} \\
0&=& {\bf C}({\bf X},\partial_i{\bf X})\label{eq:constraints},
\eeq
where eqs.(\ref{eq:evol}) are the evolution equations and eqs. (\ref{eq:constraints}) are the constraints. 
An interesting observation is that apart from the evolution equations
for $r_{\alpha}$, the spatial derivatives appear linearly. Moreover,
the only equations involving second-order spatial derivatives are
those of the Hubble gradient, $r_{\alpha}$, which contain terms of the
form $\partial_i\partial_jr_{\alpha}$. 

\subsection{The plane waves}
In our case we will be interested in a particular set of exact
solutions, namely the vacuum plane-wave solutions. In general the
plane-wave solutions have a 2-dimensional Abelian group acting freely
on the spatial hypersurfaces\footnote{Cosmological models which have a 2-dimensional Abelian group acting freely
on the spatial hypersurfaces are usually referred to as $G_2$-models \cite{WE}.}; i.e., they are special exact $G_2$ solutions to the equation of motion. In particular, by choosing the vector $r_{\alpha}$ appropriately, the $G_2$ models have $r_{\alpha}=(r_1,0,0)$ (see \cite{WCthesis} for details). The exact $G_2$ plane-waves solutions can easily be found; however, we will only consider a special class of them here. The class we are considering is the \emph{spatially homogeneous} plane-wave solutions of Bianchi type VII$_h$. These plane-wave solutions are given by 
\beq
\d s^2=e^{2\tau}(-\d \tau^2+\d x^2)+e^{2s(x+\tau)}\left[e^{\beta}(\cos v\d y-\sin v \d z)^2+e^{-\beta}(\sin v\d y+\cos v \d z)^2\right], \nonumber \eeq
where $v=b(x+\tau)$ and $b$, $\beta$ and $s$ are constants satisfying $b^2\sinh^2\beta = s(1-s)$, $0<s<1$.
The Killing vectors $\partial_y$ and $\partial_z$ span the wave-fronts
and are the Killing vectors that survive the generalisation to the
$G_2$ models. We will see later that the critical unstable modes
correspond to inhomogeneous perturbations in the $(y,z)$-plane so we
do not expect the specialisation to spatially plane-wave homogeneous
models to be critical for our result. 
 
\section{Perturbing the plane wave}
The spatially homogeneous plane waves of Bianchi type VII$_h$ are given in the expansion-normalised variables by
(the notation used is the same as in \cite{CH,HHC})
\beq 
&& q=2\chi, ~\Sigma_+=-\chi,~A_{\alpha}=A\delta^1_{\alpha}, ~A=1-\chi, ~ \Sigma_-^2=N_{23}^2=\chi(1-\chi), \\ \nonumber 
&& \bar{N}=\lambda N_{23}, ~ \text{where}~ 0<\chi<1,~ \lambda^2>1, ~(1-\chi)=3h(\lambda^2-1)\chi \\ \nonumber 
&& \Sigma_{23}=\Sigma_{12}=\Sigma_{13}=N_-=N_{11}=N_{12}=N_{13}=r_{\alpha}=\Omega=0. 
\eeq
In the following we will assume that $\Omega=0$ (vacuum)\footnote{The assumption $\Omega=0$ is not crucial for our analysis since we find that the models are unstable. Including matter will only bring more degrees of freedom, and hence more perturbative modes, into the analysis.}. Moreover, we will also chose the following gauge: 
\beq 
R_1=\sqrt{3}\lambda\Sigma_-, \quad R_2=-\sqrt{3}\Sigma_{13}, \quad R_3=\sqrt{3}\Sigma_{12}. 
\eeq
An alternative choice of gauge could be, for example, one where we eliminate some of the expansion-normalised variables; however, the above choice is useful because it simplifies the equations of motion significantly.  
We will also assume that the Codazzi equations has been used to eliminate the variables $\Sigma_{12}$ and $\Sigma_{13}$. 

\subsection{Showing instability} 
Let us first establish the fact that these plane waves are unstable to inhomogeneous perturbations. 
Consider the variable $N_{11}$, which vanishes for the background plane-wave spacetime. For the above choice of gauge the evolution equation for $N_{11}$ is 
\beq
\dt N_{11}=(q-4\Sigma_+)N_{11}.
\label{eq:N11}\eeq
At the linearised level, we get $\dt N_{11}=6\chi N_{11}$, and hence this variable is unstable. So for models that allow for a non-zero $N_{11}$, the plane waves are unstable. The constraints governing the $N_{\alpha\beta}$ variables, are the Jacobi constraint equations: 
\beq
0=\left(\dbold_{\beta}-r_{\beta}\right)\left(N^{\alpha\beta}+\eps^{\alpha\beta\gamma}A_{\gamma}\right)-2A_{\beta}N^{\alpha\beta}.
\eeq
It is clear that the general inhomogeneous models allow for a non-zero $N_{11}$; hence, the plane-wave spacetime is unstable to generic inhomogeneous perturbations. 

The component $N_{11}$ is the 1-component of the commutator between $\dbold_2$ and $\dbold_3$:
\beq
N_{11}=\left(\left[\dbold_2,\dbold_3\right]\right)_1,
\eeq
which means that a pertubation in $N_{11}$ stems from a spatial perturbation of the vectors $\dbold_2$ and $\dbold_3$. Thus to understand where this instability comes from, it is of importance to study the differential operators $\dbold_2$ and $\dbold_3$, and the Hubble gradients $r_{\alpha}=-(1/H)\dbold_{\alpha}H$. Note  that in the Abelian case (i.e., the $G_2$ models) $N_{11}=0$. 

\subsection{Interpretation} 
The frame variables, $E^{~i}_{\alpha}$, decouple for the spatially homogeneous models where ${\bf e}_0$ is orthogonal to the type VII$_h$ hypersurfaces. For the background type VII$_h$ plane waves, the frame variables can be taken to have the form 
\beq
E^{~i}_{1}&=&(1,0,0)\nonumber \\
E^{~i}_{2}&=&(0,E^{~y}_2,E^{~z}_2)\nonumber \\
E^{~i}_{3}&=&(0,E^{~y}_3,E^{~z}_3),
\eeq
where the matrix $(E^{~i}_\alpha)$ is invertible. Introducing the complex frame variable  \[ Z^i=E^{~i}_2+\frac{1}{\lambda}\left(1+ i\sqrt{\lambda^2-1}\right)E^{~i}_3, ~i=y,z,\] 
the evolution equations and the commutator relations, 
 \beq
\dt E_{\alpha}^{~i}&=& \left(q\delta^{\beta}_{~\alpha}-\Sigma^{\beta}_{~\alpha}+\eps^{\beta}_{~\alpha\gamma}R^{\gamma}\right)E^{~i}_\beta \\
0&=& 2\left(\dbold_{[\alpha}-r_{[\alpha}-A_{[\alpha}\right)E_{\beta]}^{~i}-\varepsilon_{\alpha\beta\delta}N^{\delta\gamma}E_{\gamma}^{~i}, 
\eeq
imply that for the plane waves
\beq
\dt Z^i& = &\left(3\chi-i\omega\right)Z^i, \nonumber \\
\partial_x Z^i &=& \left(A-i\omega\right)Z^i,
\eeq
where $\omega\equiv \sqrt{3\chi(1-\chi)(\lambda^2-1)}$. 
Hence, $Z^i\propto \exp[3\chi t+Ax-i\omega (t+x)]$. This implies that
the frame variables diverge for the plane waves in question. Note that
for a SH cosmology, we have that $\partial_i {\bf X}\equiv 0$, and thus $\dbold_{\alpha}{\bf X}\equiv 0$. So in these models the frame variables decouple from the equations of motion. However, as we perturb these models inhomogeneously, $\dbold_{\alpha}{\bf X}$ is no longer necessarily zero and the divergence of the frame variables may become important. The above shows that the differential operators $\dbold_2$ and $\dbold_3$ carry a divergent term $e^{3\chi t}$ which, in turn, indicates that the plane waves are unstable with regard to inhomogeneous perturbations in the $\dbold_2$ and $\dbold_3$ directions (orthogonal to the propagation of the wave). 

In the following analysis we will choose $Z^y=\exp[3\chi t+Ax-i\omega (t+x)]\equiv Z$, and $Z^z=iZ$. It is also advantageous to introduce the following complex variable, 
\[ {\sf r}=r_2+\frac{1}{\lambda}\left(1+ i\sqrt{\lambda^2-1}\right)r_3 \]
along with the operators
\beq
{\mbold \partial}& \equiv &Z^i{\partial}_i=Z(\partial_y+i\partial_z), \nonumber \\
\bar{\mbold \partial}& \equiv &\bar{Z}^i{\partial}_i=\bar{Z}(\partial_y-i\partial_z).
\eeq

To investigate the (in)stability of these spacetimes with respect to
general inhomogeneous perturbations we will linearise the equations of
motion around the exact plane-wave solutions (assuming that terms like $({\bf X}-{\bf X}_0)$, $\dbold_{\alpha}{\bf X}$, $\dbold_{\alpha}\dbold_{\beta}{\bf X}$, etc. are small). The interesting modes stem from the Hubble gradients $r_1$ and ${\sf r}$, so to understand what is happening we will linearise these equations explicitly. The evolution equation for $r_{\alpha}$ and the vorticity constraint read 
\beq
\dt r_{\alpha}&=& \left(q\delta^{\beta}_{~\alpha}-\Sigma^{\beta}_{~\alpha}+\eps^{\beta}_{~\alpha\gamma}R^{\gamma}\right)r_\beta +\dbold_{\alpha}q \\ 
0&=& \left[\varepsilon^{\alpha\beta\gamma}(\dbold_{\beta}-A_{\beta})-N^{\alpha\gamma}\right]r_{\gamma}.
\eeq
The deceleration parameter, $q$, for vacuum is given by
\[  q=2\Sigma^2+\frac 23 A^{\alpha}r_{\alpha}-\frac 13\dbold_{\alpha}r^{\alpha}. \]
From the linearised vorticity constraint, $\dbold\bar{\sf r}=\bar\dbold {\sf r}$, and using $[\dbold,\bar{\dbold}]=0$, we can write 
\beq
\dbold q=2\dbold\Sigma^2+\frac 13(2A\dbold r_1-\dbold \partial_x r_1)-\frac 13(\dbold_2^2+\dbold_3^2){\sf r}.
\eeq
The vorticity constraint also gives the following linear relation:
\beq
\dbold r_1=(\partial_x-A){\sf r}+M_{\sf r}{\sf r}+M_{\bar{\sf r}}\bar{\sf r},
\eeq
where $M_{\sf r}$ and $M_{\bar{\sf r}}$ are constants. However, for simplicity, we will leave this constraint unsolved and rather impose it on the initial conditions. 

At this stage it is useful to introduce an orthonormal set of functions, $f_{\bf k}(y,z)$, on the flat torus spanned by the vectors $\partial_y$ and $\partial_z$. These functions can be chosen to be proportional to the usual Fourier modes $e^{-i(k_yy+k_zz)}$, and hence, ${\mbold\partial} f_{\bf k}=-i{\sf k}Zf_{\bf k}$, where ${\sf k}=k_y+ik_z$. Defining 
\beq
~~\kappa_{2}&=&E^{~i}_2k_i=-\frac{i\lambda}{2\sqrt{\lambda^2-1}}\left({\sf k}Z-\bar{\sf k}\bar{Z}\right), \\
~~\kappa_{3}&=&E^{~i}_3k_i=\frac{i}{2\sqrt{\lambda^2-1}}\left[\left(1-i\sqrt{\lambda^2-1}\right){\sf k}Z-\left(1+i\sqrt{\lambda^2-1}\right)\bar{\sf k}\bar{Z}\right],
\eeq
and $\kappa^2\equiv\kappa_2^2+\kappa_3^2\geq 0$,
 we obtain
\beq 
(\dbold_2^2+\dbold_3^2)f_{\bf k}=-\kappa^2f_{\bf k}.
\eeq
Here, $\kappa^2$ is a quadratic monomial in $Z$ and $\bar{Z}$, and therefore diverges as $\kappa^2\propto e^{6\chi t}$. By expanding all the linearised variables as 
\beq
{\bf X}=\sum_{\bf k}{\bf X}_{\bf k}(t,x)f_{\bf k}(y,z), 
\eeq 
we obtain 
\beq
\dt {\sf r}_{\bf k}&=&\left(3\chi-i\omega+\frac 13\kappa^2\right){\sf r}_{\bf k}-i{\sf k}Z\left[4(\Sigma_{\alpha\beta})_0\Sigma_{\bf k}^{\alpha\beta}+\frac 23(1-\chi)r_{1,{\bf k}}-\frac 13\partial_xr_{1,{\bf k}}\right].\nonumber \\
\eeq

We note that we can write 
\beq
(|{\sf r}_{\bf k}|^2)'=\left(6\chi+\frac 23\kappa^2\right)|{\sf r}_{\bf k}|^2-2\Real(i{\sf k}Z\rho_{\bf k}\bar{\sf r}_{\bf k}),
\eeq
where $\rho_{\bf k}$ is a linear combination of $\Sigma_{\bf k}^{\alpha\beta}$, $r_{1,{\bf k}}$ and $\partial_xr_{1,{\bf k}}$. For ${\bf k}\neq 0$, there exist $\left\{\kappa^2_{\text{min}}(x),\kappa^2_{\text{max}}(x)\right\}>0$ such that  
\[ e^{6\chi t}\kappa^2_{\text{min}}(x)\leq \kappa^2\leq e^{6\chi t}\kappa^2_{\text{max}}(x).\] 
Let us also introduce 
\beq 
I\equiv \frac 23\int_{t_0}^t\kappa^2\d t,
\eeq 
which obeys the bounds
\beq 
 \frac{\kappa^2_{\text{min}}(x)}{9\chi}(e^{6\chi t}-e^{6\chi t_0})\leq
 I\leq\frac{\kappa^2_{\text{max}}(x)}{9\chi}(e^{6\chi t}-e^{6\chi t_0}).\nonumber 
\eeq
The equation for $r_1$ has a similar $\kappa^2$ term, so taking into account the vorticity constraint, we would naively expect that  $|{\sf r}_{\bf k}|^2\propto e^{\lambda t+I}$ for a constant $\lambda$, and hence, ${\sf r}_{\bf k}$ is diverging. 

We are thus led to conclude that the plane waves are unstable with respect to inhomogeneous perturbations of the Hubble gradient ${\sf r}$. 

\section{Discussion}
In this paper we have shown that the plane-wave spacetimes of Bianchi type VII$_h$ are unstable with respect to inhomogeneous perturbations. The cause of this instability ultimately stems from the behaviour of the frame variables $E_{\alpha}^{~i}$, which are unbounded into the future for the plane-wave solutions. For the spatially homogeneous solutions these frame variables decouple from the equations of motion; however, in the general $G_0$ models this unboundedness of the frame variables is crucial. As we pointed out earlier, the expansion-normalised differential operators $\dbold_{\alpha}$ carry this divergence and are thus in some sense intrinsically unstable. In particular, we showed that the Hubble gradient, $r_{\alpha}=-(1/H)\dbold_{\alpha}H$, has two unbounded components corresponding to directions orthogonal to the propagation of the wave. 

We  also point out that the limit $\chi\rightarrow 0$ corresponds to
the vacuum Milne model. The above analysis has excluded this case
since the eigenvalue $(3\chi-i\omega)$ gets a troublesome zero real
part. In this case a separate analysis is required. Pertubations of
the Milne universe have been studied in \cite{Bruni} which shows that this
model is also unstable. 

For the $G_2$ models, $r_{\alpha}=(r_1,0,0)$ and hence the  critical
unstable modes obtained here are excluded.  A non-zero $r_1$ is still
allowed, and from the equation of motion we note that there is a 'zero
eigenvalue' corresponding to the evolution equation for $r_1$. This is
really no surprise since we know that there exist more general plane
wave solutions with $r_1\neq 0$. It is not known what role these $G_2$ plane waves play in the evolution of general $G_2$ cosmologies. 

Another point worth mentioning is the choice of temporal gauge. In this work we used the separable volume gauge, defined in eq.(\ref{temporalgauge}); however, another useful gauge is the synchonous gauge. Also in this gauge we can show that the plane waves are unstable; in particular, eq.(\ref{eq:N11}) remains on the same form. This shows the robustness of our result with regards to the temporal gauge. 

Recently, Barrow and Tsagas \cite{BT} have studied linearised
perturbations of the same plane-wave spacetimes. However, they
considered the subcase in which the Weyl tensor does not vary along the
directions $\partial_{y}$ and $\partial_z$. This assumption alone
effectively reduces the model to a $G_2$ model and thus, as we pointed
out eariler, the unstable critical mode we found here is not included. 

In the analysis the unboundedness of the frame variables $E_{\alpha}^{~i}$ was critical for the unstable nature of these models. We may thus ask ourselves: Will all models which have unbounded frame variables $E_{\alpha}^{~i}$ be unstable against generic inhomogeneous perturbations? Future research may shed some light on this issue. 

\section*{Acknowledgments} 
The authors would like to thank John D. Barrow, Henk van Elst, Woei Chet Lim and Christos Tsagas for useful discussions and comments.  
This work was funded by a Killam PostDoctoral Fellowship (SH) and NSERC (AC). 

\appendix

\bibliographystyle{amsplain}

\end{document}